# `stdchk`: A Checkpoint Storage System for Desktop Grid Computing

Samer Al Kiswany [*], Matei Ripeanu [*], Sudharshan S. Vazhkudai [†], Abdullah Gharaibeh [*]
[*] The University of British Columbia {samera, matei, abdullah}@ece.ubc.ca
[†] Oak Ridge National Laboratory vazhkudaiss@ornl.gov

*Abstract*— Checkpointing is an indispensable technique to provide fault tolerance for long-running high-throughput applications like those running on desktop grids. This paper argues that a dedicated checkpoint storage system, optimized to operate in these environments, can offer multiple benefits: reduce the load on a traditional file system, offer high-performance through specialization, and, finally, optimize data management by taking into account checkpoint application semantics. Such a storage system can present a unifying abstraction to checkpoint operations, while hiding the fact that there are no dedicated resources to store the checkpoint data.

We prototype `stdchk`, a checkpoint storage system that uses scavenged disk space from participating desktops to build a low-cost storage system, offering a traditional file system interface for easy integration with applications. This paper presents the `stdchk` architecture, key performance optimizations, support for incremental checkpointing, and increased data availability. Our evaluation confirms that the `stdchk` approach is viable in a desktop grid setting and offers a low-cost storage system with desirable performance characteristics: high write throughput and reduced storage space and network effort to save checkpoint images.

## I. INTRODUCTION

Checkpointing is an indispensable fault tolerance technique adopted by long-running applications. These applications periodically write large volumes of snapshot data to persistent storage, in an attempt to capture their current state. In the event of a failure, applications recover by rolling-back their execution state to a previously saved checkpoint.

The checkpoint operation and the associated data have unique characteristics. First, applications have distinct phases where they compute and checkpoint; often, these phases occur at regular intervals. Second, checkpointing is a write I/O intensive operation. Consider a job running on thousands of compute nodes: this scenario has the potential to generate thousands of files, amounting to terabytes of snapshot data at each timestep. Under these conditions, high-resolution checkpointing can easily overwhelm the I/O system. Third, checkpoint data is often written once and read only in case of failure. This suggests that checkpoint images are seldom accessed beyond the lifetime of an application run or even during the run. Finally, checkpointing, however critical it may be for reliability, is pure overhead from an application standpoint, as time is spent away from useful computation. To minimize this overhead, expensive high-throughput storage devices are often used.

This paper argues that the above characteristics of the checkpoint operation can be exploited to design a high-performance, yet low-cost, storage system, optimized to serve checkpointing needs. As a proof of concept we focus on a desktop grid scenario but our architecture can be applied to any storage system integrating a large number of unreliable components (e.g., a cluster). Desktop grids are collections of loosely connected machines—within a single administrative domain—harnessed together to provide compute cycles for high throughput applications. Several academic and industry solutions support this scenario and current deployments aggregate thousands of nodes [1]. In a desktop grid, checkpointing stresses the storage system further as it is not only used to increase application reliability but also to support process migration when a user reclaims a machine.

This article presents a checkpoint-optimized storage system that aggregates storage contributions from nodes participating in a desktop grid. A checkpoint-optimized storage system can bring significant benefits. First, such a system offloads the I/O intensive checkpoint operations from the traditional file system, thus alleviating the load on an expensive shared server. Second, this storage system can be optimized for the checkpoint operation; for example, it can reduce file system overhead associated with large writes as well as reduce data storage requirements. As a result applications can checkpoint at a significantly higher rate than what is currently feasible with shared file systems. Third, checkpoint data is transient in nature and is often not maintained beyond the lifetime of a successful application run. Unlike a regular file system, a checkpoint storage system can be aware of this characteristic and act like a cache to purge or prune files using a combination of data usage, aging and user specified policies. Finally, a checkpoint storage system needs to present only a unifying file system abstraction and can hide the fact that there are no dedicated resources to store the checkpoint data. Further, the storage system can even be built atop unreliable resources (storage donors), much like how a computational desktop grid itself is based on an unreliable substrate (cycle donors).

*Our contributions* in this paper are as follows. We present `stdchk`, a checkpoint storage system for HPC applications in a desktop grid environment. Much like how `stdin` and `stdout` input/output systems are ubiquitously available to applications, we argue that checkpointing is an intense I/O operation, requiring a special 'data path'. We show that this data path can be made available to HPC applications as a dedicated, low-cost checkpoint-optimized storage system.

While we base our solution on our previous storage system work that aggregates scavenged disk space to build a cache (write-once-read-many semantic) [2], `stdchk` is optimized for a different workload namely, *high-speed writes of incremental versions of the same file*. To this end, `stdchk` introduces several optimizations to render itself 'checkpoint-friendly' to HPC applications:

- *High sequential write throughput.* `stdchk` exploits the I/O parallelism that exists inherently in a desktop grid to provide a suite of optimized write protocols that enable checkpointing at throughputs higher than what is feasible in current desktop grid settings. Our results indicate an observed application bandwidth of 110MB/sec in a LAN connected desktop grid.
- *Support for incremental versioning.* `stdchk` minimizes the size of the data stored using a novel solution to incremental checkpointing that exploits the commonality between successive checkpoint images. We put forth several heuristics that do not require application or operating system support to identify commonality between incremental versions of the same checkpoint image. We evaluate these heuristics in the context of real applications. Our results indicate a substantial reduction in checkpointed data size and generated network traffic. A desired side-effect of this approach is that it enables applications to checkpoint at a finer granularity.
- *Tunable data availability and durability.* Since `stdchk` aggregates storage contributions from transient workstations, standard replication techniques are used to ensure data availability and durability. Further, applications can decide the level of availability/durability they require.
- *Tunable write semantics.* Additionally, `stdchk` gives applications the ability to choose between a write semantic that is pessimistic (return only after the desired level of replication is achieved) or optimistic (return immediately after data has been written safely once, while replication occurs in the background). This further gives applications control over the write throughput vs. data durability tradeoff.
- *Automatic pruning of checkpoint images.* `stdchk` offers efficient space management and automatic pruning of checkpoint images. These data management strategies lay the foundation for efficient handling of transient data.
- *Easy integration with applications:* `stdchk` provides a traditional file system interface, using FUSE (File system in User SpacE) Linux kernel module [3], for easy integration with applications.

To summarize, the novelty of our approach lies in recognizing that checkpointing can benefit from specialized storage system support and in bringing to bear a combination of storage solutions to address this specific workload. Our system builds on best practices from several domains: storage scavenging from peer-to-peer storage systems, striping from parallel I/O, incremental checkpointing from versioning storage, data archival, data availability and durability from replicated storage and applies them to checkpointing.

The rest of this paper is organized as follows: the next section surveys the related work, section III presents the design considerations for a checkpointing oriented storage system, section IV details `stdchk` design, while section V presents an extensive evaluation study. We conclude in section VI.

II. RELATED WORK

To the best of our knowledge, there is no dedicated storage system for checkpointing either in a desktop grid or in a large-scale cluster environment. However, the following efforts are relevant to this paper.

*Support for checkpointing.* Applications can use one of the following checkpointing solutions, offering different transparency vs. performance tradeoffs. Application-level checkpointing offers no transparency, yet performance is high as the application programmer manages the process of collecting and saving the application state. Checkpointing libraries (e.g., BLCR [4], DejaVu [5]), often available on large-scale clusters, provide an increased level of transparency by checkpointing at the process level. However, checkpointing libraries do not provide full transparency as application programmers still have to manually insert checkpointing calls into the application code. Finally, system-level checkpointing offers complete transparency at the cost of ignoring application semantics, which can often be used to reduce storage requirements. None of these techniques entail storage system support and simply use the file system as is.

*Workload-optimized storage systems.* Building storage systems geared for a particular class of I/O operations or for a specific access pattern is not uncommon. For example, the Google file system [6] optimizes for large datasets and append access; the Log-structured file system [7] optimizes for writes, arguing that most reads are served by ever increasing memory caches; BAD-FS [8] optimizes for batch job submission patterns; while FreeLoader [2] optimizes for write-once read-many data access and exploits the locality of interest and sequential access patterns of scientific data analysis. Parallel file systems (Lustre [9], PVFS [10], GPFS [11]) also target large datasets and provide high I/O throughput for parallel applications. In a similar vein, a checkpoint optimized storage system can be geared toward this write intensive HPC I/O operation.

*Relaxing POSIX semantics.* Another related thread is the relaxation of POSIX semantics so that the file system can better cater to HPC I/O. In this vein, Lightweight File System [12] provides a small set of critical functionality that an I/O library can extend to build solutions for parallel applications. For `stdchk`, we have chosen to provide a POSIX interface for easy integration with applications, while at the same time building an high-performance storage system.

*Contributory storage.* A number of storage systems [2, 13-15] aggregate space contributions from collaborating users to provide a shared data store. Their base premise is the availability of a large amount of idle disk space on personal desktops that are online for the vast majority of the time. The specific technical solutions, however, vary widely as a result of different targeted deployment environments (local vs. wide-area networks), different workloads (e.g., unrestricted file-system workload vs. read-only workload vs. checkpointing workload), or different assumptions on user motivation to contribute storage (e.g., from systems that propose storage space bartering to motivate users to systems that assume collaborative users by default).

*Versioned Storage.* Several file systems save periodic snapshots of an entire file system to recover from accidental file deletions or changes. Examples include Plan 9 [16] or AFS [17] that use a single policy that guides file retention for the entire file system and the Elephant file system [18] that incorporates user-specified policies to determine the versions to retain. On the one side, the checkpoint scenario is more coarse-grained in that each checkpoint is written sequentially. The flip side is that copy-on-write techniques used by the aforementioned systems offer no gains when entire files are written sequentially.

*Low bandwidth file systems.* To reduce the amount of data sent to remote storage, LBFS [19] detects similarity between file versions sent over the network by only transmitting the changed file ranges. This is similar to utilities such as CVS that transmit deltas of files to bring server and user copies up to date. We evaluate LBFS techniques in our setting and find that their overhead is too high.

**Relationship with authors' previous work** on storage aggregation. Our previous work includes the FreeLoader project [2], a read-only storage system for large datasets that aggregates scavenged space. It demonstrates that scavenged storage can be efficiently aggregated from LAN-connected nodes to provide a fast read-only data cache. It further explores striping techniques' impact on read throughput and load on contributory nodes. Vazhkudai et al. [2] demonstrate a peak read throughput of 88MB/s from a stripe width of ten 100Mb/s benefactors, using round-robin striping.

We have chosen to build `stdchk` by making use of two concepts from this previous work: storage aggregation using scavenging and striping. Based on these concepts, we have built a sophisticated storage infrastructure geared towards distributed checkpointing. The design and implementation of the storage system has been fundamentally modified to incorporate new functionality conducive to checkpointing namely: file versioning, file replication, garbage collection, session semantics, and optimized write techniques that enable delegating to applications, the control of the tradeoffs between data reliability and performance.

### III. DESIGN CONSIDERATIONS FOR A CHECKPOINT STORAGE SYSTEM

Applications running in a desktop grid have the following options for storing checkpoint images.

- **Node-local storage.** It is common practice for jobs running on the individual nodes in a desktop grid to checkpoint to their respective node-local storage. Local storage is dedicated and is not subject to the vagaries of network file system I/O traffic. Moreover, local I/O on even moderately endowed desktops offers around 30-50MB/sec. However, local storage is bound to the volatility of the desktop itself. First, individual nodes in a desktop grid might have to be relinquished as soon as the owner returns, leaving little time to migrate checkpoint data. Second, desktops are themselves not highly reliable and failure is common. Thus, the checkpoint data, saved on local storage, is lost when the node crashes.
- **Shared file systems.** Alternatively, nodes within a desktop grid can also checkpoint to a shared, central file server. However, shared file systems are usually crowded with I/O requests and have limited space. Further, the hundreds of nodes in the desktop grid— on which processes of a parallel application may be running—can flood the central server with simultaneous checkpointing I/O operations.
- **Distributed checkpointing.** A desirable alternative, adopted by this paper, is the construction of a distributed storage system optimized for a checkpointing workload. Such a targeted storage infrastructure can be built by aggregating storage contributions from the individual nodes within the desktop grid itself—much like how CPU cycles are aggregated to form the desktop grid.

#### A. Checkpoint I/O Workload Characteristics

This section summarizes the characteristics of a typical checkpoint workload in a desktop grid.
- *Applications typically create one file per process per timestep.* Thus, a large parallel application, running for a few hours and checkpointing every 15 minutes, can easily create tens of thousands of files.
- *Applications have distinct compute and checkpoint phases.* Parallel applications on thousands of nodes can simultaneously access the storage system to save their images.
- *Checkpoint data is transient in nature.* Successive checkpoint images are produced throughout an application's lifetime. These images are accessed only in the case of a failure, process migration, or for debugging or speculative execution. Depending on the usage scenario, a checkpoint image may become obsolete at the end of a checkpoint interval when a new image is produced, after the successful execution of the application, or, in case of migration, at process restart time. Alternatively, a checkpoint image might be useful in the long term for debugging and speculative execution.
- *Low risk.* Checkpoint image loss involves rolling-back the computation to the image corresponding to the previous timestep. While, in the common case, this may affect the job turnaround time, data loss effects are dependent on the specific application execution scenario.

With these workload characteristics in mind, let us look at some design goals for a checkpoint storage system.

*B. Design Goals*

This section describes the desirable properties of a checkpoint storage system.
- *Performance.* Elmootazbellah et al. [20] identify the performance of the storage system as the key driver for checkpointing overheads. Consequently, the checkpoint storage should be optimized for write performance, while a reasonable read performance is necessary to support timely job restarts.
- *Easy-to-use interfaces.* The storage system should provide an interface that enables easy integration with applications. Specialized libraries and interfaces, however optimized, cannot match the simplicity of file system interfaces.
- *Low overhead.* Although file system interfaces are desirable their overhead should be minimal. For instance, the overhead involved in metadata management and synchronization, can all be minimized for checkpoint storage.
- *Support for incremental checkpoints and data sharing.* To reduce the storage and I/O load, the storage system should be able to exploit data commonality between successive checkpoints.
- *Scalability.* The storage system should scale to support a large number of simultaneous client requests. For instance, multiple nodes, on which a parallel application is running, will likely checkpoint all at once.
- *Flexible namespace.* The storage system should provide a flexible naming scheme that enables easy identification of an entire set of checkpoint images as belonging to a particular application's checkpoint operation. Additionally the namespace should support versioning.
- *Support for checkpoint image management.* The storage system should include components to manage checkpoint image lifetime according to user specified policies: e.g., complete replacement of checkpoint images when a new image is produced, removal of all images at successful completion of application, or long-term storage of checkpoint images.

IV. STDCHK: A CHECKPOINT-FRIENDLY STORAGE SYSTEM

Storage space scavenging is a good base for building a low-cost storage system in desktop grids as parallelism in these environments is achieved by exploiting infrastructure that is already in place rather than using expensive hardware.

*A. System Architecture*

**Overview.** `stdchk` integrates two types of components: a *metadata manager* and a number of *benefactor* (or donor) nodes that contribute their free storage to the system. Datasets are fragmented into smaller *chunks* that are striped across benefactor nodes for fast storage and retrieval. This basic model is common to other storage systems (e.g., GoogleFS, FreeLoader) as well.
- *The metadata manager* maintains the entire system metadata (e.g., donor node status, file chunk distribution, and dataset attributes). Similar to a number of other storage systems we have chosen a centralized metadata manager implementation.
- *The benefactor nodes* contribute their storage space to the system. To facilitate integration, our design minimizes the set of functions storage nodes provide: they interact with the manager to publish their status (on-/off-line) and free space using soft-state registration, serve client requests to store/retrieve data chunks, and run garbage collection algorithms.

Data storage and retrieval operations are initiated by the client via the manager. To retrieve a file the client first contacts the metadata manager to obtain the chunk-map, (i.e., the location of all chunks corresponding to the file), then, the actual transfer of data chunks occurs directly between the storage nodes and the client.

When a new file is written, `stdchk` cannot predict in advance the file size. Thus, storage space allocation is done incrementally. Clients eagerly reserve space with the manager for future writes. If this space is not used, it is asynchronously garbage collected.

The manager also stores metadata regarding benefactor space contributions, file versioning and replication as we describe in this section. Reads and writes are performed in parallel to a *stripe width* of benefactors in chunks of fixed size, using a round-robin striping policy. The storage system is particularly geared for high-speed writes. To this end, `stdchk` offers a suite of write protocols and innovative storage system support for incremental checkpointing. In addition, `stdchk` offers tunable background replication and write semantics.

The storage system is mounted under `/stdchk`. Any file opened under this mounting directory is written to the aggregate storage system, thereby making `stdchk` easily available to client applications. The rest of this section describes `stdchk`'s main design choices.

***Session Semantics.*** A key decision shaping the design of a distributed storage system is the consistency model. Existing systems differ widely in terms of their write consistency semantics. Solutions range from unspecified consistency semantics (e.g., NFS [21]) to strong consistency, provided, for example through access serialization [22]. Our storage system provides session semantics [23]. Data commits are delegated to `stdchk` client proxies: when, the client application eventually performs a *close()* operation, the client proxy will commit the chunk-map for the dataset to the manager. The fact that this operation is atomic ensures session consistency. We note that, strictly speaking, session semantic is not necessary for checkpointing operations as checkpoint images are immutable and have a single producer. However, introducing a clear and low-overhead consistency model gives a good path for future transitioning of `stdchk` towards a generic high-performance file system.

***Dealing with failures: Reliable writes.*** Since `stdchk` stores data in a distributed storage cloud, failure of donor nodes needs to be addressed to ensure reliable operation. Nodes can fail after the chunks have been stored on them. This issue is addressed by replicating data over multiple storage nodes. However, replication introduces a new question: should a write operation return immediately after the first replica of the data has been persistently stored or wait until all data reaches the desired replication level. The tradeoff is between data-loss risk and write throughput. A client can choose to be conservative (pessimistic) and wait until a desired level of replication is achieved before declaring a write operation successful. In this case, the client favors data durability over high write throughput. Alternatively, an optimistic client can return as soon as a chunk is written to the first benefactor and let the background replication process bring about the desired replication level. In this case, the client favors write throughput over data durability. The choice between optimistic and pessimistic writes is a client configuration parameter.

***Data replication: User-defined replication targets.*** In our target environment, storage nodes are unreliable desktops. Any solution, addressing data availability, needs to factor the following: (1) facilitate fast writes so the application can quickly return to performing useful computation, (2) reliably store checkpoint data so that it is available if needed, and (3) provide good read performance to minimize restart delays.

To this end, we evaluated both erasure coding and replication. Erasure coding incurs significant computational overhead compared to replication. The checkpointing application has to compute the erasure code while writing the data. Alternatively, if this computation is performed in the background, after the write, it leads to significant network traffic to pull the different chunks to a single node, perform the encoding and distribute them again. Further, data reads involve equivalent computational and network traffic overheads.

Replication, on the other hand, incurs no computational overhead, but involves larger space overhead for the same degree of reliability. Replication can be implemented as a background task, thereby imposing minimally on the application. Further, replication is easier to implement as it involves less complex data management. Finally, since checkpoint data is mostly transient in nature, the space overhead is transient. In some cases, the application might choose to keep the images for a prolonged duration, in which case, the data can be offloaded to more stable storage.

For these reasons, we have chosen replication to improve data reliability. Replication is implemented as a background task initiated by the manager. The manager builds a *shadow-chunk-map* for the checkpoint dataset that comprises of a list of benefactors to host the replicas of the original chunks. The process of building a shadow-map is similar to the process of selecting a stripe width of benefactors for a new write operation. The shadow-map is then sent to the source benefactors to initiate a copy to the new set of benefactors. Once the copy succeeds the shadow-map is committed to the manager. Creation of new files has priority over replication so that applications' writes can be expedited.

The other case of failure is the manager node failure. The manager can fail before the client has had a chance to push the final chunk-map of the dataset. This results in the manager metadata being inconsistent with the current state of the benefactors. To address this case, we can extend the client functionality to push the final chunk-map after the write, to benefactors. The benefactors can then update the manager (once back online) with metadata about the new file. Once the manager has received concurrence from two-thirds of the width of benefactors, it can safely add metadata about the new checkpoint dataset. A hot-standby manager as a failover is another option in such cases.

***Garbage collection.*** To decouple, to the extent possible, benefactors from metadata management, the deletion operation is performed only at the manager which results in orphaned chunks at benefactors. To reclaim space, benefactors periodically send a list of the set of chunks they store and the manager replies with the set of chunks that can be garbage collected.

*B. Write Optimizations for High Throughput*

Our implementation optimizes large, sequential writes, the most frequent operation in a checkpointing storage system. Depending on whether local I/O, remote I/O or a combination of the two is used, and based on the overlap of these operations, four designs are possible:

- **Complete local write.** When enough local space is available, the write operation can temporarily dump the application data to the local node. Then, it can asynchronously push it out to the `stdchk` when the application closes the file. The main advantage of this approach is its simple implementation. The drawback, however, is that it makes intense use of the local disk to store data that is eventually pushed out of the local node. Further, checkpointing to local storage does not protect against local node failures.
- **Incremental write.** The node-local storage may not always have enough free space to temporarily store the entire checkpoint image. Even when space is available, it is not reliable. Incremental writes limit the size of the temporary file. When the temporary file size reaches a predefined limit, writes are redirected to a new temporary file, and the previous one is pushed to `stdchk`. Once all incremental files have been pushed out and after the application has issued a *close()*, the chunk-map for the complete file is pushed to the metadata manager. While this solution still uses local I/O, it overlaps data creation and remote propagation, leading to faster overall transfer to remote storage.
- **Sliding window write.** To minimize the application perceived write delay, we exploit the fact that, for most modern desktops, the disk write bandwidth is lower than the achievable network throughput achievable with commonly used Gigabit NIC. The *sliding window* technique pushes data out of the write memory buffer directly to `stdchk` storage. This method completely eliminates the use of local disk.

*C. Support for Incremental Checkpointing*

A checkpoint image typically involves a dump of the application's memory, comprising data structures and other state variables. Incremental versions of the same application image may produce (partially) similar files. This property can be used to improve the write throughput and/or reduce storage and network requirements, ultimately providing support for checkpointing at higher frequency. The challenge, however, is to detect similarities at runtime without operating system or application support.

To investigate whether the similarity between checkpoint images can indeed be exploited in real settings we address the following three interrelated issues. First, we evaluate the potential gains from detecting similarity between successive checkpoint images. Second, we evaluate heuristics to understand to what degree the task of detecting file similarity can be efficiently implemented by the storage system without application or operating system support. Third, we design the architecture to efficiently support these heuristics.

*Heuristics to detect similarities.* The generic problem of identifying the maximum common substring between two strings has computational and space overheads unacceptable in the context of file systems. We thus evaluate two heuristics that offer lower overheads.

- *Fixed-size compare-by-hash (FsCH):* This approach divides the checkpoint image into equal-sized chunks, hashes them and uses the hashes to detect similar chunks. The main weakness of this approach, however, is that it is not resilient to file insertions and deletions. An insertion of only one byte at the beginning of an image prevents this technique from detecting any similarity.
- *Content-based Compare-by-hash (CbCH):* Instead of dividing the file into equal-sized blocks, CbCH detects block boundaries based on file content (as suggested by LBFS [19]). CbCH scans the file using a 'window' of $m$ bytes and, for each position of the window, computes a hash of the corresponding string. A chunk boundary is declared if the lowest $k$ bits of the hash are all zero. Then, identification of chunk similarities proceeds as above based on chunk hashes. Statistically, $k$, the number of bits of the hash compared to zero allows controlling the average chunk size, while $m$, the window size, and $p$, the number of bytes the window is advanced every time, allow controlling the variation in chunk sizes and additionally influence the chunk size. Unlike FsCH, CbCH is resilient to data insertion/deletion, since inserting/deleting some bytes will only affect one block (two blocks if the changes are at block boundary). The drawback is that CbCH requires hashing more data, hence results in larger computational overhead.

Section V.E includes an extensive performance evaluation of these heuristics using two real-world applications. We evaluate the rate of similarity detected and the computational overhead for application-/library-/VM-level checkpointing and different checkpoint intervals. Our results suggest that FsCH is the best approach for `stdchk` due to the balance it offers between throughput and reduced space consumption as a result of similarity detection.

*Architectural support.* To support these heuristics and to manage incremental checkpoint images efficiently, `stdchk` provides the following:

- *Content based addressability.* `stdchk` provides content-based naming of data chunks, that is, to name chunks based on a hash of their content. An additional advantage of using content-based naming is that it enables data integrity checks. This feature can be used to prevent faulty or malicious storage nodes from tampering with the chunks they store.
- **Support for copy-on-write and versioning**. Additionally, stdchk supports versioning and copy-on-write, so that chunks that have been identified as similar can be shared between different file versions. When a new version of a checkpoint image is produced, only the new chunks need to be propagated to persistent storage. The new chunk-map will integrate the newly produced chunks and the chunks that have already been stored.

### D. Support for Automated, Time-Sensitive Data Management

The burden of managing large volumes of data (checkpoint or output data) that HPC applications produce can become onerous. We aim to build into the storage system, the intelligence to automatically manage files based on user-specified policies concerning their lifetimes. To this end, stdchk exploits the fact that checkpoint images are often used in a few standard scenarios. Most of the checkpoint data is time sensitive. For example, in a normal application scenario, checkpoint images are made obsolete by newer ones; while in a debugging scenario, all checkpoint images may need to be saved to enable debugging.

We support this functionality through versioning, the use of a simple naming convention that helps recognize successive files from the same application, and the integration of user-specified metadata. By convention, files in stdchk are named as follows: $A.N_i.T_j$ stands for an application $A$, running on node, $N_i$ and checkpointing at timestep $T_j$. We treat all images—from the many processes of application $A$ running on nodes, $N$—as versions of the same file. Files from an application are organized within a folder for that application. The folder has special metadata concerning the time-related management of the files that it contains. Currently we support the following scenarios:
- *No intervention*: All versions (from multiple timesteps) are persistently stored indefinitely.
- *Automated replace*: New checkpoint images make older ones obsolete.
- *Automated purge*: Checkpoint images are automatically purged after a predefined time interval.

### E. Providing a Traditional File System Interface

The strong requirement for a file system-like interface is motivated by two observations. First, a traditional file-system interface is crucial for adoption and increased usability of the storage system. Second, in the specific context of checkpointing systems, the libraries that support checkpointing are complex pieces of code that, in some situations, are executed in kernel mode. Modification or even recompilation to integrate them with a custom storage system would be a high barrier to adoption and may be considered a security risk.

A number of implementation alternatives are possible to provide a file-system interface. One approach is to build a Virtual File System (VFS) module [24]. The Linux kernel provides hooks for adding new file systems via loadable modules without requiring kernel recompilation. NFS [21], for example, is implemented using this approach. While this approach provides good performance, custom kernel modules are considered a security risk in some of our target deployment environments.

We adopt a user–space file system implementation for three additional reasons. First, we can avoid the complexity and maintenance overhead of kernel-level development. Second, using a module that handles all the system call details allows the developer to focus on storage system logic rather on system calls and VFS details. Finally, it is possible to hide the extra context switch overhead by overlapping local I/O operations and the actual data transfer. Several projects have adopted this approach with reasonable overhead (e.g., Ceph [25]). This is also confirmed by our experience.

To implement a user-space file system, one option is to use an interception technique to catch file-system calls, filter those related to checkpointing, and redirect them to a user-level implementation. Parrot [26] and early versions of PVFS [10] adopted this approach. However, this approach results in high overhead to maintain the interception code along with the evolution of system-calls [10].

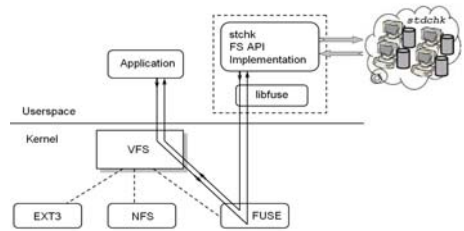

**Figure 1**. File system call path through FUSE.

We use FUSE, a Linux kernel module[3], similar to the other VFS modules (e.g. NFS, ext3). Once a FUSE volume is mounted, all system calls targeting the mount point are forwarded to the FUSE kernel module, which preprocesses and forwards them to user-level file system callbacks (see Figure 1). When the callback function finishes processing the system call, FUSE post-processes the call and returns the results to VFS. FUSE is officially merged into the Linux kernel starting with 2.6.14 version, further simplifying adoption of our user-space file system.

Our user-space file system implementation maps the system calls to stdchk operations. Additionally, it handles granularity differences. For example, applications usually write in small blocks, while remote storage is more efficiently accessed in data chunks of the order of a megabyte. Further, our implementation is performance optimized for our deployment scenario. It provides high-performance writes, improves read performance through read-ahead and high volume caching, and caches metadata information so that most system *readdir* and *getattr* system calls can be answered without contacting the manager.

## V. EVALUATION

We evaluate our prototype under a range of micro- and macro-benchmarks. Except where specifically mentioned, we use a testbed composed of 28 machines. Each machine has two 3.0GHz Xeon processors, 1 GB RAM, two 36.5G SCSI disks, and Gigabit Ethernet cards. For all configurations we report (using error bars in plots) averages and standard deviations over 20 runs.

### A. Platform Characterization

We first evaluate the performance and the overhead of each individual component. The sustained write throughput achievable on a local disk with write caches enabled is 86.2MB/s, while accessing a dedicated NFS server deployed on the same node achieves 24.8MB/s.

We also use micro-benchmarks to estimate the overhead due to the additional context switch any user-level file system entails. Thus, to evaluate FUSE module overheads we have built two simple file systems. The first one (FUSE to local I/O in Table I) simply redirects all write requests back to the local file system. The second (/stdchk/null) ignores the write operation and returns control immediately. Table 1, presents the time to write a 1 GB file to the local disk and to these two file systems. The results show that FUSE overhead is very low, about 2%, on top of local I/O operations. The /stdchk/null performance indirectly indicates that the cost of the additional context switch using FUSE is about 32μs.

**Table 1** Time to write a 1 GB file.

|  | Local I/O | FUSE to local I/O | /stdchk/null |
|---|---|---|---|
| Average Time (s) | 11.80 | 12.00 | 1.04 |
| Standard deviation | 0.16 | 0.24 | 0.03 |

### B. Write Throughput

Our write implementation decouples the application write I/O from the actual file transfer over the network that stores the file on donor nodes. Therefore, we define two performance metrics to compare the various alternatives for write-optimized operations described in Section IV.B). First, *the observed application bandwidth* (OAB) is the write bandwidth observed by the application: the file size divided by the time interval between the application-level *open()* and *close()* system calls. Second, *the achieved storage bandwidth* (ASB) uses the time interval between file open() and until the file is stored safely in stdchk storage (i.e., all remote I/O operations have completed).

Figure 2, presents the OAB when the number of remote nodes to save data on (the stripe width) varies from one to eight benefactors. Our experiments show that two contributing nodes with 1 Gbps NICs can saturate a client. However, when benefactors are connected by a lower link bandwidth (100Mbps), a larger stripe width is required to saturate a client (our technical report presents these experiments [28]).

The *complete-local-write* OAB is similar to that of FUSE local writes. This is not surprising since all data is written to a local temporary file and then transferred to storage nodes after the file close operation.

Higher concurrency allows *sliding window and incremental writes* to perform better in terms of OAB (at around 110 MB/s). This high bandwidth translates to shorter time for checkpoint operation as observed by the application. Further, sliding window interface completely avoids the local IO and hence its performance is mainly influenced by the amount of memory buffers allocated to the interface. Section V.C evaluates this effect.

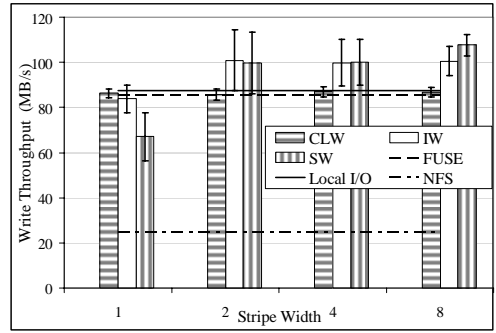

**Figure 2.** The average observed application bandwidth (OAB) for three write optimized techniques: complete local writes (CLW), incremental writes (IW), and sliding window (SW). For comparison the figure also shows: the throughput of writing to the Local-I/O, to local I/O through the FUSE module (FUSE), and to a dedicated NFS server (NFS) running on the same node.

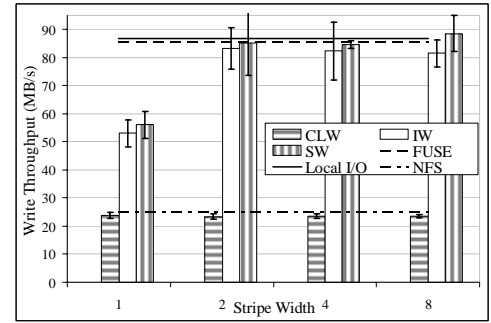

**Figure 3.** The average achieved storage bandwidth (ASB) for complete local writes (CLW), incremental writes (IW), and sliding window (SW).

Figure 3, presents the *achieved storage bandwidth* (ASB). Since complete-local-write serializes the local write and network transmission, it performs worst.

Incremental and sliding-window write interfaces achieve better concurrency. The performance of the complete local write improves only slightly when adding more benefactors since the bottleneck is the local I/O. Sliding-window performs best and, in our experiments, we saturate the Gigabit network card with only two benefactors. Figures 2 and 3 also show that sliding-window write performance is slightly better than either local I/O or network file system based checkpointing. Even though the local I/O in this case is comparable to our write interfaces, data stored on local nodes is volatile due to the transient nature of contributed storage.

*C. Effect of Configuration Parameters*

The size of the temporary file and the size of the memory buffers used have a significant impact on the write performance. This section investigates this impact. Figure 4 and Figure 5 present the observed and achieved throughput for the sliding window write with different allocated memory buffers. As shown in the figures, sliding window is able to saturate the network link with only two benefactors.

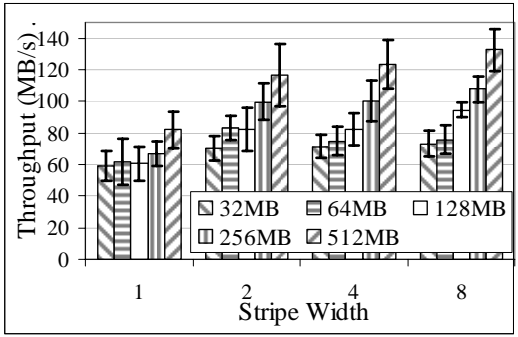

**Figure 4.** The observed application bandwidth (OAB) for the sliding window write for different number of benefactors and allocated buffer size (in MB).

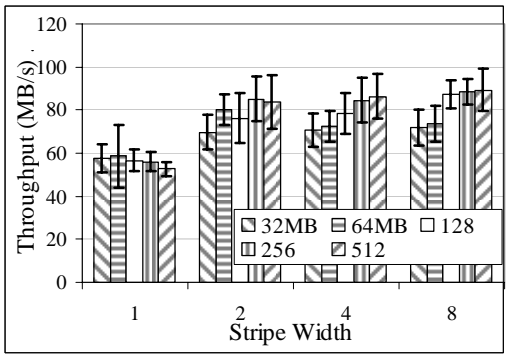

**Figure 5.** The achieved storage bandwidth (ASB) for sliding window writes for different number of benefactors and allocated buffer sizes (in MB).

Similarly the incremental write interface performance is affected by the size of the temporary files. Our experiments indicate that smaller temporary files result in larger OAB and ASB due to higher concurrency in the write operation. Due to space constraints we do not present this result.

*D. Write Performance on a 10Gbps Testbed*

We have tested $stdchk$'s sliding window write performance on a testbed endowed with higher IO, network and processing capabilities. The testbed is composed of a single $stdchk$ client and four benefactor nodes. The client is a Xeon 2GHz processor, 8GB memory, SATA disk, and a 10Gbps NIC. The benefactors have Xeon 1.6 GHz processor, 8GB memory, SATA disks, and 1Gbps NIC.

Figure 6, presents the OAB and ASB of the sliding window write interface. The interface's buffer size is set to 512MB. As Figure 6 shows, with four benefactor nodes, $stdchk$ is able to successfully aggregate the I/O bandwidth of the contributing benefactors and achieves up to 325 MB/s of OAB and 225 MB/s of ASB. (We note that the scale of this experiment is limited by the size of the testbed we currently have access to)

This experiment shows that $stdchk$ can efficiently integrate multiple benefactors to provide a high write throughput as the number of benefactors nodes grows.

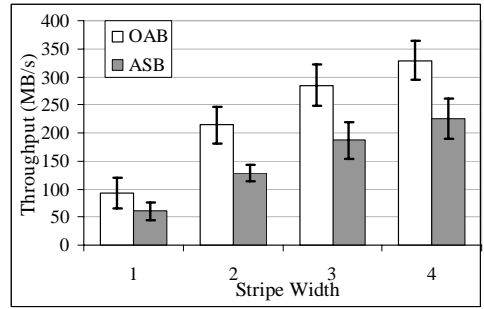

**Figure 6.** The observed application bandwidth (OAB) and achieved storage bandwidth (ASB) of the sliding-window interface with varying the stripe width.

*E. Incremental Checkpointing: Feasibility and Performance*

This section presents evidence that supports our decision to include incremental checkpointing in $stdchk$. We evaluate the potential gains from detecting similarity between successive checkpoint images and the performance of heuristics that operate at the file system level without application or operating system support. Additionally we evaluate the performance of our complete storage system implementation.

The two heuristics we compare (described in section IV.C): fixed-size compare-by-hash (FsCH) and content-based compare-by-hash (CbCH) differ in their efficiency of detecting similarities and in the imposed computational overhead.

To quantitatively evaluate these heuristics along these two axes and to ground our comparison in the real-world, we use checkpoint-images from two popular scientific applications: a protein-substrate complex biomolecular simulation (which we call BMS [29] for brevity), and BLAST [30], a bioinformatics protein/nucleic acid sequence searching tool. BMS uses application-level checkpointing and we have instrumented BLAST with library (using BLCR [31]) and virtual machine-based checkpointing (using Xen [27]). Table 2 presents the trace details.

**Table 2:** Characteristics of the collected checkpoints.

| Application | Checkpointing type | Checkpoint interval (min) | # of checkpoints | Average checkpoint size (MB) |
|---|---|---|---|---|
| BMS | Application | 1 | 100 | 2.7 |
| BLAST | Library (BLCR) | 5 | 902 | 279.6 |
| BLAST | Library (BLCR) | 15 | 654 | 308.1 |
| BLAST | VM (Xen) | 5 | 100 | 1024.8 |
| BLAST | VM (Xen) | 15 | 300 | 1024.8 |

Table 3 presents the average ratio of the detected similarity and the achieved throughput (in MB/s) for the two techniques we compare. For each technique, the table presents the performance for key parameterization points (our accompanying technical report presents a more detailed study). The results show that, in general:

- *We can not detect similarity with application-level checkpointing*. This is due to the user-controlled, ideally-compressed format used to create these checkpoint images.
- *The level of similarity for system-level checkpointing techniques is extremely high*. For example, BLAST, using library based checkpointing (BLCR), generates checkpoints with up to 84% average similarity between successive images.

A surprising result is the near-zero similarity observed using virtual machine based checkpointing. We have verified that this is due to the particular way in which Xen checkpoints. Xen optimizes for speed, and when creating checkpoints it saves memory pages in essentially random order. Further, to preserve the ability to recreate correct VM-images, Xen adds additional information to each saved memory page. We are currently exploring solutions to create Xen checkpoint images that preserve the similarity between incremental checkpoint images.

**Table 3:** Performance comparison of similarity detection heuristics. The table presents the average rate of detected similarity and the throughput in MB/s (in brackets) for each heuristic.

| Technique | | BMS | BLAST | | |
|---|---|---|---|---|---|
| | | *App* | *BLCR* | | *Xen* |
| | | *1 min* | *5 min* | *15 min* | *5 or 15 min* |
| *FsCH* | 1KB | 0.0% [96] | 25% [99] | 9% [100] | Low similarity for both FsCH and CbCH techniques. |
| | 256KB | 0.0% [102] | 24.3% [110] | 7.1% [112] | |
| | 1MB | 0.0% [108] | 23.4% [109] | 6.3% [113] | |
| *CbCH* | overlap | 0.0% [1.5] | 84% [1.1] | 70.9% [1.1] | |
| | no-overlap $m$=20B, $k$=14b | 0.0% [28.4] | 82% [26.6] | 70% [26.4] | |

From Table 3 we further observe that:
- FsCH has the highest throughput but pays in terms of similarity detection between successive checkpoints.
- With CbCH, aggressively configured to detect block boundaries, the similarity rate is extremely high. However, this significantly reduces the achievable throughput. When the window to detect block boundaries is advanced by one byte every time (labeled 'overlap' in Table 3), throughput degrades to as low as 1MB/s. Advancing the window with its size every time (labeled 'no-overlap' in Table 3), can improve throughput to about 26MB/s, which is still four times slower than FsCH.

The CbCH results thus far present only an upper-bound for similarity detection but do not explore the tradeoff between similarity detection, throughput, and block size. We explore this tradeoff in Table 4 and present the effect of varying $m$ (the window size) and $k$ (the number of bits compared to zero to detect a block boundary) on the CbCH no-overlap performance. We use the BLAST/BLCR trace with 5-minute checkpoint intervals. In general, as the window size $m$ increases, the ratio of detected similarity decreases, mainly due to the reduced opportunity to detect block boundaries, leading to larger blocks. On the other hand, we can control the block size by varying the number of zero bits we require to detect a boundary: lower $k$ leads to smaller blocks. However, as $k$ increases the variation in the block size increases (the table presents averages for the minimum and maximum detected block for each checkpoint image).

**Table 4:** The effect of $m$ and $k$ on CbCH no-overlap performance. The table presents the ratio of detected similarity (in percentage), the heuristic's throughput in MB/s, the average resulted checkpoint size in KB, and the average minimum and maximum chunk sizes (Values for $m$ in bytes and for $k$ in bits)

| $k$ | $m \rightarrow$ | *20* | *32* | *64* | *128* | *256* |
|---|---|---|---|---|---|---|
| **8** | *Similarity (%)* | *30.0* | *62.8* | *62.4* | *64.3* | *64.5* |
| | Throughput (MB/s) | 85.7 | 86.8 | 86.3 | 86.0 | 84.2 |
| | Avg. size (KB) | 519.2 | 522.4 | 530.7 | 547.3 | 579.5 |
| | Avg. min size (KB) | 325.1 | 275.6 | 210.1 | 350.2 | 257.1 |
| | Avg. max size (KB) | 614.3 | 627.3 | 668.9 | 787.3 | 967.9 |
| **10** | *Similarity (%)* | *38.6* | *72.4* | *66.3* | *65.0* | *64.7* |
| | Throughput (MB/s) | 75.6 | 78.2 | 77.5 | 74.6 | 69.5 |
| | Avg. size (KB) | 539.3 | 552.5 | 584.7 | 654.8 | 778.9 |
| | Avg. min size (KB) | 265.9 | 283.9 | 294.7 | 409.2 | 380.8 |
| | Avg. max size (KB) | 893.9 | 890.0 | 1095.0 | 1491.2 | 2251.7 |
| **12** | *Similarity (%)* | *77.3* | *73.4* | *65.6* | *63.0* | *60.7* |
| | Throughput (MB/s) | 47.0 | 53.6 | 50.2 | 52.3 | 53.6 |
| | Avg. size (KB) | 626.3 | 665.4 | 812.5 | 1076.3 | 1544 |
| | Avg. min size (KB) | 239.8 | 242.2 | 269.5 | 437.7 | 456.2 |
| | Avg. max size (KB) | 1683.8 | 1807.8 | 2632.5 | 3812.7 | 4510.4 |
| **14** | *Similarity (%)* | *82.4* | *71.7* | *61.3* | *58.4* | *57.1* |
| | Throughput (MB/s) | 26.6 | 32.7 | 34.2 | 40.6 | 46.43 |

| Avg. size (KB) | 930.8 | 1079.2 | 1635.6 | 2267.3 | 2908.6 |
|---|---|---|---|---|---|
| Avg. min size (KB) | 514.9 | 232.0 | 449.5 | 528.8 | 506.8 |
| Avg. max size (KB) | 3710.9 | 3639.5 | 4515.1 | 4662.2 | 4646.6 |

For the `stdchk` prototype, we have chosen to integrate FsCH as it offers higher throughput and a simpler implementation path. We are currently exploring alternatives to provide a high-performance CbCH implementation by offloading the intensive hashing computations to the Graphical Processing Unit (GPU). As the `stdchk` write throughput is the main success metric, we chose to implement FsCH in `stdchk`. FsCH detects reasonable similarity with highest data throughput. The other techniques are prohibitively slow in detecting commonality between checkpoints.

Figure 7 presents the sliding window write's average OAB and ASB with and without FsCH. (labeled SW-FsCH and SW-no-FsCH respectively in the figure). The test involved writing 75 successive checkpoint images of BLAST using BLCR. The sliding window write is tested with different memory buffer sizes and with four benefactors. The checkpointing interval is 5 minutes, the chunk size is 1MB and the average checkpoint size is 280MB. The figure shows a slightly degraded write performance with SW-FsCH (116MB/s average OAB and 84 MB/s ASB). The main advantage, however, is the reduced storage space and network effort (by 24%). One exception to this result is when the write interface is configured with large memory buffers (256MB). In this case, the overhead for detecting similarities leads to 25% lower OAB (but similar ASB). This is explained by the small checkpoint image size (280 MB on average), which allows storing nearly all the data in the write buffer and makes the OAB performance solely dependent on *memcopy* performance (similarity detection is slowed down by the hashing overhead).

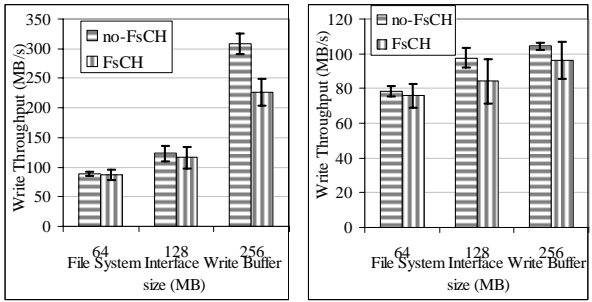

**Figure 7.** The observed application bandwidth (OAB, left plot) and the achieved storage bandwidth (ASB, right plot) for the sliding window with and without incremental checkpointing supported by fixed block compare-by-hash.

### F. `stdchk` Scalability

To assess the scalability and performance under heavy load, we start the `stdchk` system with 20 benefactors and 7 clients on the 28 node testbed. Each client writes 100 files of 100MB each. This workload translates to around 70 GB of data, and 2800 manager transactions (four for each write operation). To ramp-up the load, clients start at 10s interval. Figure 8, presents the aggregate throughput of the storage system. We observe a sustained peak throughput of about 280MB/s limited, in fact, by the networking configuration of our testbed. This demonstrates that our system is able to scale to match an intense workload.

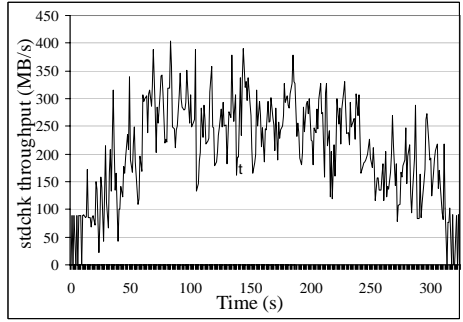

**Figure 8.** `stdchk` throughput at larger scale: 7 clients generate a synthetic workload to stress a `stdchk` pool supported by 20 benefactor nodes.

### G. Putting Everything Together

To complement our synthetic performance benchmarks we have used so far, and to evaluate `stdchk` performance when integrated in a full system we compare the performance of the BLAST application when checkpointing on the local disk and on `stdchk`. The application is configured to checkpointing every 30 The `stdchk` testbed uses four benefactors with 1Gbps network connections. Table 5, presents the execution time as well as the total amount of checkpoint data generated for both runs. The improvement in overall execution time is minor (1.3%) as the overall ratio of application execution time to checkpointing time is high. More importantly, the results presented in Table 5, show that `stdchk` speeds up the checkpointing operation itself by 27% and leads to a 69% reduction in storage space and network effort.

**Table 5,** The execution time and volume of generated data for BLAST application checkpointed to local disk and `stdchk`.

| | Local disk | `stdchk` | Improvement |
|---|---|---|---|
| Total execution time (s) | 462,141 | 455,894 | 1.3% |
| Checkpointing time (s) | 22,733 | 16,497 | 27.0% |
| Data size (TB) | 3.55 | 1.14 | 69.0% |

### VI. SUMMARY

This paper presents the design and implementation of `stdchk`, a distributed checkpoint storage system targeting a desktop grid environment. We have put forth arguments that support the premise that checkpointing I/O is a write intensive operation, requiring novel solutions. `stdchk` aggregates storage space from LAN connected desktop machines to provide a traditional file system interface that facilitates easy integration with applications. `stdchk` offers several checkpoint-specific optimizations such as a suite of write protocols for high-speed checkpoint I/O, support for checkpoint data reliability, incremental checkpointing and lifetime management of checkpoint images. Our prototype evaluation indicates that `stdchk` can offer an application perceived checkpoint I/O throughput as high as 110MB/sec, which is significantly higher than what is feasible with current local I/O or network file system based checkpointing. Our novel solution to exploit similarity between incremental checkpoint images results in significantly lower storage space and network effort requirements. In summary, our experience indicates that a checkpoint storage based on space aggregation is a viable solution for desktop grid applications.

## VII. Acknowledgements

This research is sponsored by grants from the Laboratory Directed Research and Development Program of Oak Ridge National Laboratory (ORNL), managed by UT-Battelle, LLC for the U. S. Department of Energy under Contract No. DE-AC05-00OR22725 and the Natural Sciences and Engineering Research Council of Canada and Canadian Foundation for Innovation.

In addition, the authors would like to thank Stephen Neville at The University of Victoria for providing access to the cluster, Derek Church for configuring it, Pratul Agarwal at ORNL for the application-level checkpoint images, Baoqiang Yan for his work on data replication, and Elizeu Santos-Neto for insightful comments on earlier versions of this paper.